\documentclass{an}
\usepackage{graphicx,psfig,cite,epsf}
\usepackage{times}
\usepackage{subfigure}

\begin{document}

\Pagespan{1}{last} 
\Yearpublication{}
\Yearsubmission{}
\Month{} 
\Volume{} 
\Issue{} 
\DOI{} 
\title{Six years of XMM-Newton observations of NGC 1313 X-1 and X-2}
\author{\textbf{Fabio Pintore\inst{1,2} \and Luca Zampieri\inst{2}}} 
\titlerunning{Six years of XMM-Newton observations of NGC 1313 X-1 and X-2}
\authorrunning{Pintore \& Zampieri}
\institute{Dipartimento di Astronomia, Universit\`a di Padova, I-35122 Padova, Italy 
\and INAF-Osservatorio Astronomico di Padova, I-35122 Padova, Italy}
\received{} 
\accepted{} 
\publonline{} 
\keywords{accretion, accretion disks -- X-rays: binaries -- X-Rays: galaxies -- X-rays: individuals (NGC 1313 X-2)}

\abstract{We present a systematic analysis of the X-ray spectra of NGC 1313 X-1 and NGC 1313 X-2, 
using six years of XMM-Newton observations (17 observations). We fitted
the continuum with a Comptonization model plus a multicolor blackbody disc, 
that describes the effects of an accretion disc plus an optically
thick corona. We checked the consistency of this spectral model on the
basis of the variability patterns of its spectral parameters. We found
that the two sources show different spectral states. We tentatively interpret the observed behaviour
of NGC 1313 X-1 within the framework of super-Eddington accretion and
that of NGC 1313 X-2 within the framework of near Eddington accretion. 
We also attempted to determine the chemical abundances in the local
environment of NGC 1313 X-1 and X-2, analyzing the RGS spectra of the longest observation available
(122 ksec). The results appear to indicate solar metallicity for the local environment of NGC 1313 X-1
and sub-solar metallicity for NGC 1313 X-2. 
}
\maketitle

\section{Introduction}
\label{introduction}

The barred spiral galaxy NGC 1313 hosts three Ultra luminous X-ray sources (ULXs),
one of which (NGC 1313 X-3) is a known supernova (SN 1978K) interacting with the
circumstellar medium. The other two ULXs, NGC 1313 X-1 and X-2 (X-1 and X-2 hereafter),
are located in different positions, close to the nucleus X-1 and at the outskirts of the host 
galaxy X-2. NGC 1313 has been observed several times by {\it XMM-Newton}
and a sufficient number of X-ray spectra are now available to attempt a
characterization of their spectral variability. 

The first analysis of the {\it XMM-Newton} spectra of X-1 and X-2 was published
by Miller et al. (2003). They fitted the spectra with an absorbed multicolor blackbody (MCD) plus 
a powerlaw model, finding that the temperature of the MCD component is much lower
than that observed in Galactic black hole (BH) binaries. The high luminosity and cool disc component 
were interpreted as evidence for the existence of intermediate-mass black hole (IMBH) of 
100�-1000$M_\odot$ (Miller et al. 2003, 2004). Feng \& Kaaret (2005) also found
that the early {\it XMM-Newton} spectra of X-1 and X-2 were best fitted with a power-law 
plus multicolor disc blackbody model, commonly used to describe the spectra of accreting BHs.
Later, Feng \& Kaaret (2006) fitted a sequence of 12 {\it XMM-Newton} observations with the same 
spectral model and found an anti-correlation between the disc luminosity and inner disc
temperature. They concluded that the soft component does not originate in a standard accretion disc.
Also Goncalves \& Soria(2006) questioned the robustness of the cool disc spectral component
and, therefore, its use for estimating the BH mass. The optical and X-ray variability of X-2 
was also investigated by Mucciarelli et al. (2007) to costrain the properties of the system.

Finally, in an independent analysis, Stobbart et al. (2006) found that a possible curvature 
above 2-3 keV is present in high counting statistics spectra. 
More recently, adopting a more physically consistent Comptonization model and using
very long exposures, Gladstone et al. (2009) showed that many ULXs, including X-1 and X-2, 
display distinct spectral curvature above 2 keV. This is interpreted as caused by
an optically thick corona that hides the inner part of the accretion
disc, in what is likely an extreme form of the so-called very high
state of Galactic BH candidates (Done \& Kubota 2006).

In this work we try to characterize the spectral variability of X-1 and X-2.
Recently, a similar analysis was performed by Vierdayanti et al. (2010) on {\it XMM} and 
{\it Swift} data of Ho IX X-1. 

\section{Data reduction}
\label{sect1}

We re-analyzed all the available {\it XMM-Newton} spectra of the ULXs hosted in the 
spiral galaxy NGC 1313, X-1 and X-2, with homogeneous criteria. The 17 observations
span a time interval of six years, from 17 October 2000 to 16 October 2006. 
Three observations were excluded because of high flare contaminations. 
Data were reduced using SAS v. 9.0.0. EPIC-MOS and EPIC-pn spectra were extracted 
selecting the good time intervals without background contamination and with a 
background count rate not higher than 0.45 count s$^{-1}$.
We used a 35" and 30" circular extraction region for X-1 and X-2, respectively, and a 65" 
circular region for the background. 

RGS spectra were extracted using the \textit{rgsproc} task with the option \textit{spectrumbinning=lambda}. 
In this way it turns out to be possible to combine the spectra of different observations. 
The EPIC spectra were grouped with a minimum of 25 counts per channel, while RGS spectra with
50 counts per channel.

All the spectral fits were performed using XSPEC v. 12.5.1.
To improve the counting statistics, we fitted the EPIC-pn and EPIC-MOS spectra together. 
EPIC spectral fits were performed in the 0.3-10.0 keV energy range, while RGS fits were 
limited to the 0.4-2.0 keV energy range. 
Following Gladstone et al. (2009), we modelled thermal Comptonization adoting the \textit{comptt} 
model (Titarchuck 1994) or the \textit{eqpair} model (Coppi 1999). When
necessary, we added a multicolor black body disc model to the spectral fit (\textit{diskbb} in XSPEC).
The \textit{comptt} model is an analytic model describing Comptonization of soft photons, 
with a Wien input spectrum, in a hot plasma. The \textit{eqpair} model
allows for a 'hybrid' (thermal and non-thermal) plasma
and computes the resulting comptonized spectrum without assuming that the electrons
are non relativistic. The seed photons may have a disc or blackbody spectral distribution. 
For ULXs, non-thermal processes are not likely to be important, so we decided to adopt a
simplified version of \textit{eqpair}, dubbed \textit{eqtherm}, that neglects them. 

\section{Results}

We analyzed 14 out of the 17 {\it XMM-Newton} observations of X-1 and X-2 adopting
an absorbed Comptonization plus multicolor blackbody disc components.
The interstellar absorption was modelled with the \textit{tbabs} model in {\it XSPEC}. 
We fixed the Galaxy column density, along the line of sight, at $0.39\cdot 10^{21}$ cm$^{-2}$ 
(Dickey \& Lockman 1990) and added a free absorption component to model local
absorption near the source. We found that the X-1 column density is significantly variable 
during time and goes from a minimum of  $1.1\cdot 10^{21}$ cm$^{-2}$ to a maximum value 
of about $3.0\cdot 10^{21}$ cm$^{-2}$ for both \textit{comptt} and \textit{eqtherm} model. 
On the other hand, 12 out of 14 observations of X-2 have values of the column density 
clustering around $1.4\cdot 10^{21}$ cm$^{-2}$. The other two observations have 
$N_H=0.3\cdot 10^{21}$ cm$^{-2}$ and $N_H=0.5\cdot 10^{21}$ cm$^{-2}$.
If we fix the column density of these two observations equal to $1.4\cdot 10^{21}$ cm$^{-2}$,
the spectral parameters remain essentially unchanged. 

\begin{figure}[!htbp]
 \subfigure{\includegraphics[height=6.0cm,width=8.0cm]{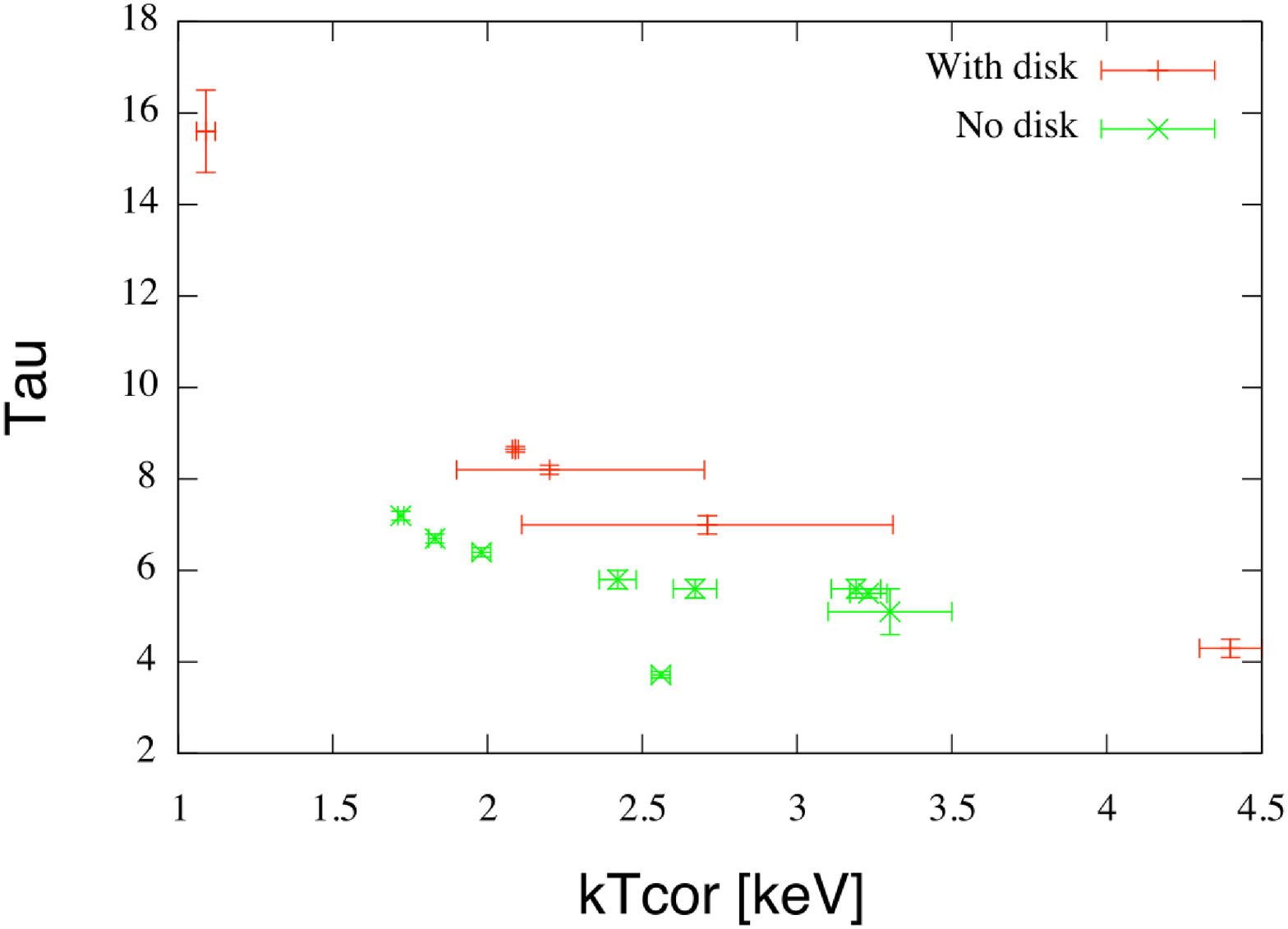}}
 \subfigure{\includegraphics[height=6.0cm,width=8.0cm]{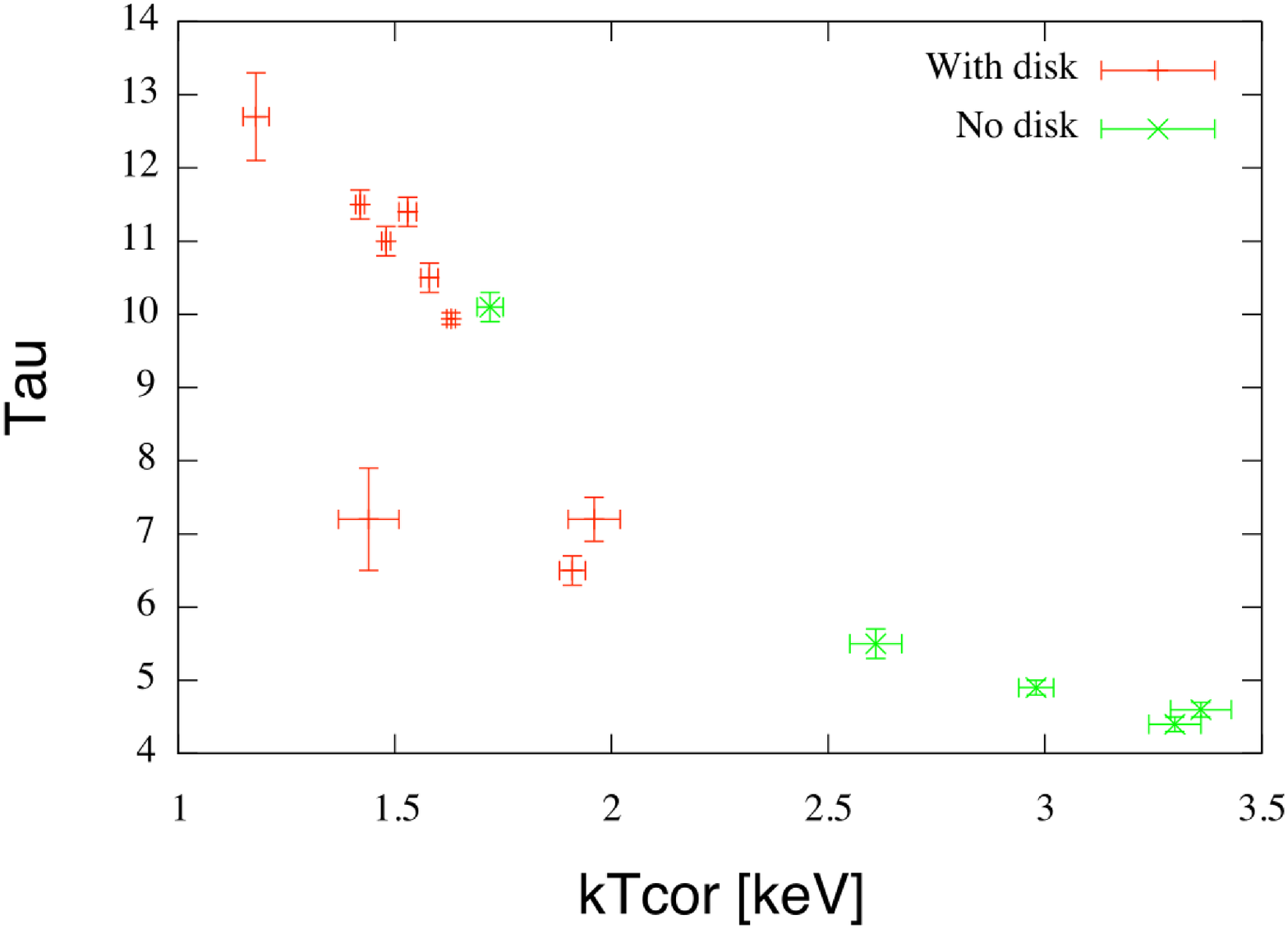}}
\caption{Optical depth versus temperature of the corona for the X-ray
spectral fits of X-1 ({\it top}) and X-2 ({\it bottom}) with a \textit{comptt}+\textit{diskbb} model.
}
\label{figts}
\end{figure}

\begin{figure}[!htbp]
\includegraphics[height=5.5cm]{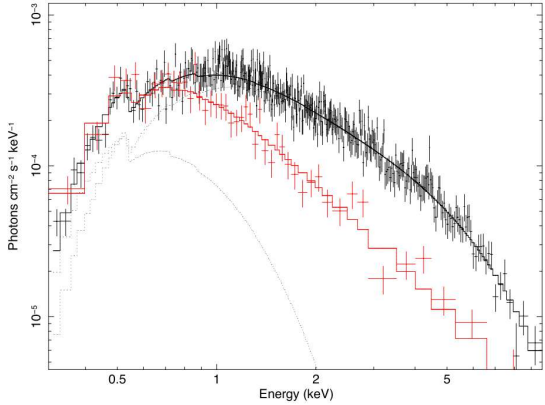}
\caption{X-ray spectra of X-2 for a low luminosity state {\it without} disc ({\it red}) 
and a high luminosity state {\it with} disc ({\it black}).}
\label{figsp}
\end{figure}

The best fitting parameters of the corona for X-1 and X-2, obtained modelling Comptonization
with \textit{comptt}, are reported in Figure~\ref{figts}. All errors
are at the 90\% confidence level for one interesting parameter.
The position of the observations of X-1 on the $kT_{cor}$-$\tau$ plot does not appear
to show a clear correlation with the presence or absence of a soft component. 
The temperature of the soft component is in the range 0.21-0.35 keV.
On the other hand X-2 appears to show a well defined behaviour in the $kT_{cor}$-$\tau$ plane
that, apart from a single observation, is correlated with the presence 
of a \textit{diskbb} component. The temperature of this component 
lies in the range 0.25-0.38 keV. When the soft component is present, the corona temperature
is low ($\la$ 2 kev) and the optical depth is high ($\tau\ga 7$). 
When the soft component is not needed, $T_{cor}\sim 3$ keV and $\tau \sim 5$.
The spectra in these two different states are shown in Figure~\ref{figsp}. 
The only 'anomalous' observation without a soft component
located in the region of high $\tau$-low $k T_{cor}$ in Figure~\ref{figts} (bottom) is one of the two with low $N_H$
mentioned above. However, fixing the column density equal to $1.4\cdot 10^{21}$ cm$^{-2}$ and adding a soft
component, the fit of this observation is statistically equivalent to that obtained applying only the 
{\it comptt} model (with fixed $N_H$),
and the normalization of the soft component is comparable to that obtained from the other spectra that
show the presence of a disc. 
Therefore, it is unclear if this observation is really anomalous 
or if there is not sufficient statistics a low energies to pinpoint the soft component.

Letting aside this observation,
when the disc component is present the total luminosity of X-2 may vary significantly, reaching 
$\sim 8 \cdot 10^{39}$ erg s$^{-1}$.
However, at low luminosity the source appears to show spectra with and without 
a soft component, likely reflecting the lack of one-to-one correspondence between
spectral shape and luminosity observed in Galactic BH binaries.

\begin{figure*}[htbp]
\begin{center}
\subfigure{\includegraphics[height=5.7cm,width=7.6cm]{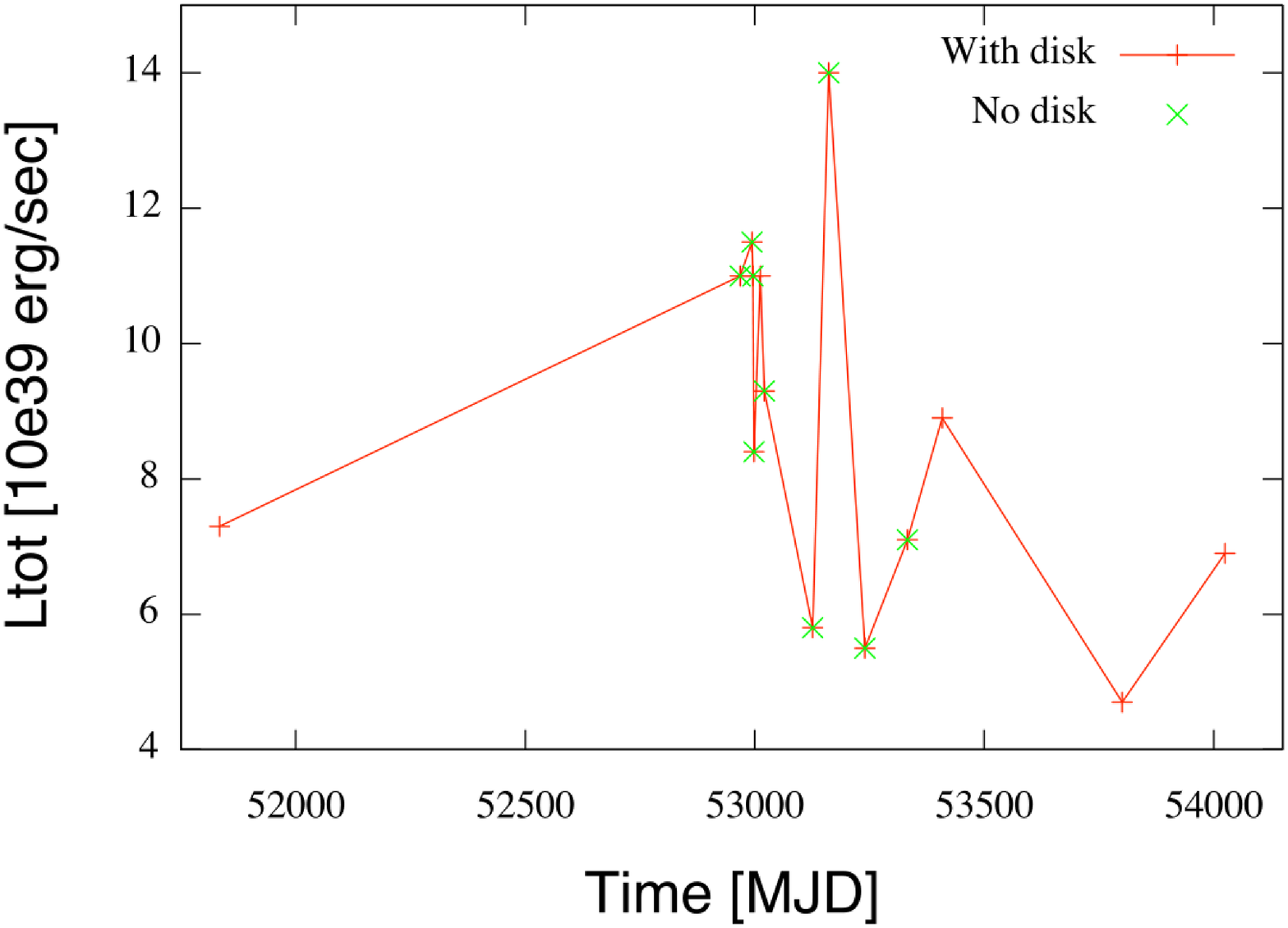}}
\subfigure{\includegraphics[height=5.7cm,width=7.6cm]{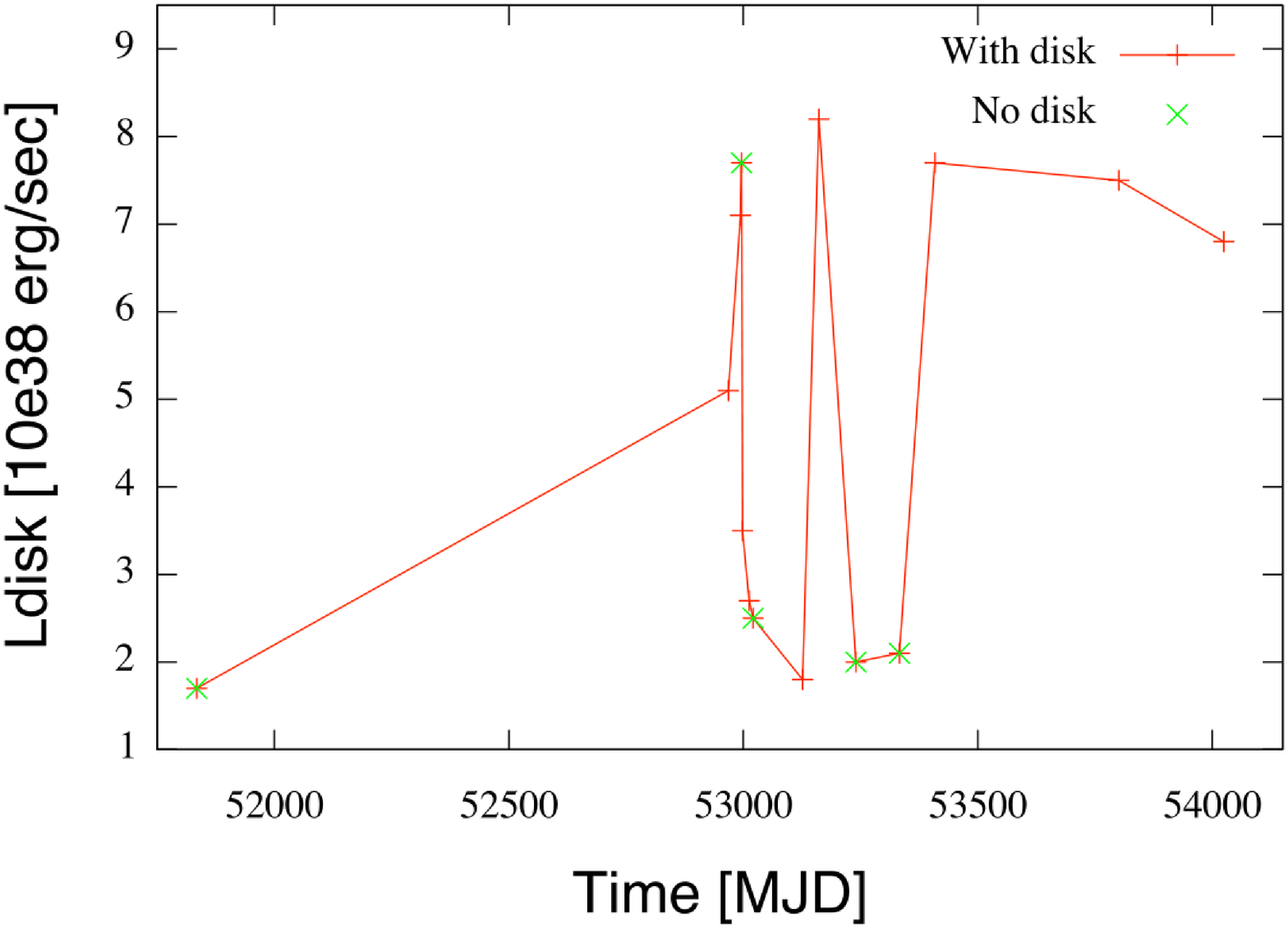}}
\caption{{\it Top}: X-ray light curve of X-1 ({\it left}) and X-2 ({\it right}).
Unabsorbed luminosities evaluated in the 0.3-10 keV energy band.}
\label{figlc}
\end{center}
\end{figure*}

Figure~\ref{figlc} shows the light curves of X-1 and X-2 computed from the
best fitting \textit{diskbb}+\textit{comptt} models (a distance of 3.7 Mpc was 
assumed; Tully 1988). On average, X-1 has a luminosity ($\sim 10^{40}$ erg s$^{-1}$),
higher than that of X-2 ($\sim 5 \cdot 10^{39}$ erg s$^{-1}$). 
Variability of a factor $\sim 3$ and $\sim 5$ is observed for X-1 and X-2, respectively.

\begin{figure}[htbp]
\begin{center}
\includegraphics [height=5.7cm,width=7.6cm]{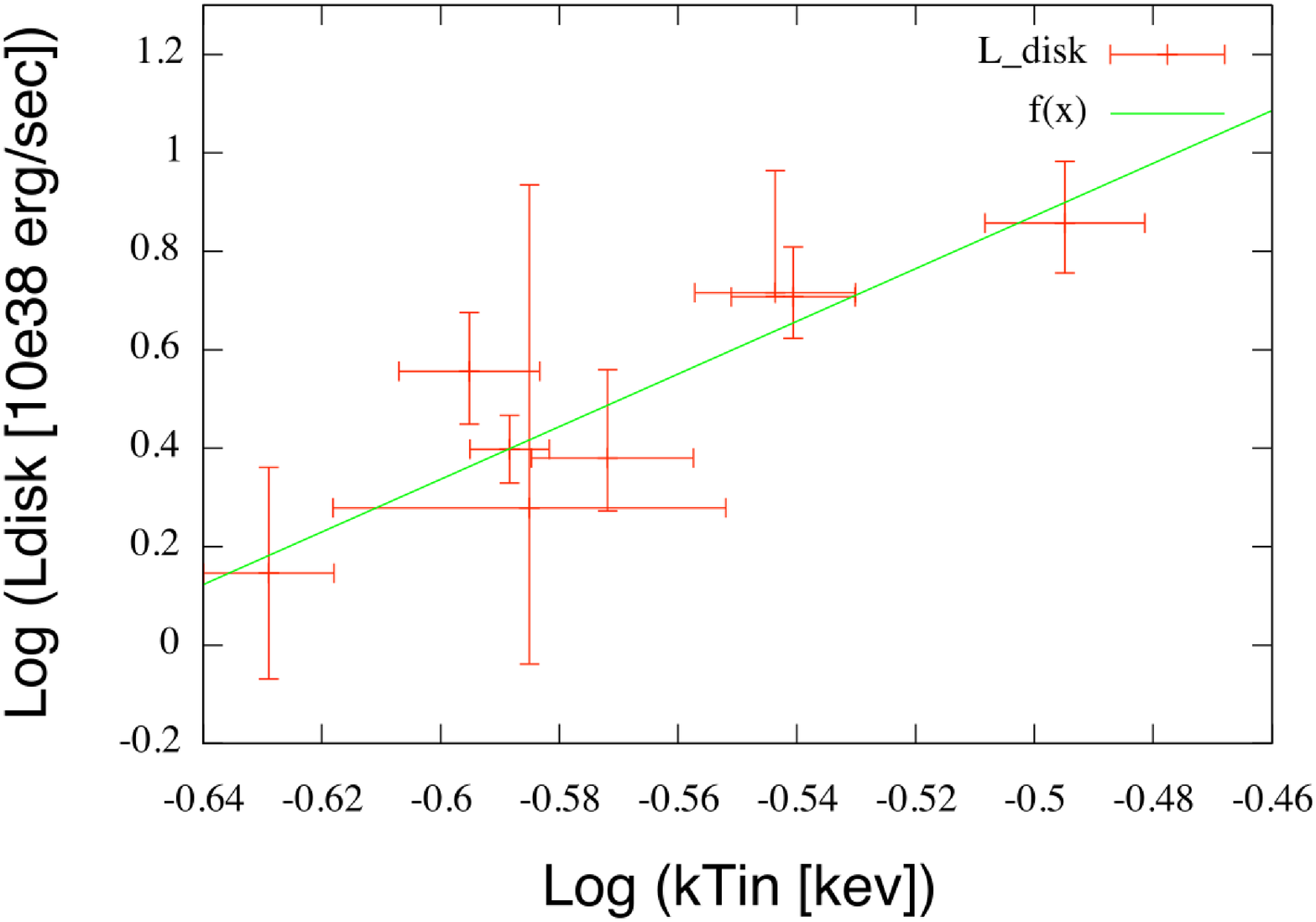}
\caption{Inner disc temperature vs disc luminosity for X-2.
The solid line is the best fit obtained with a power-law (see text). All the (unabsorbed) 
luminosities are evaluated in the 0.3-10 KeV energy band.}
\label{figlt}
\end{center}
\end{figure}

We further investigated the behaviour of the soft component in both sources. 
X-1 shows the disc component only in 5 observations while X-2 in 9 observations. 
In X-1 the luminosity of the soft component is $\sim 10^{39}$ erg s$^{-1}$ and it represents 
a significant part of the total flux. There is essentially no correlation between the luminosity
of the disc component and the inner disc temperature. The same holds true also for the total luminosity. 
On the other hand, for X-2 the luminosity of the soft component is only a small fraction 
of the total luminosity and the disc luminosity ($L_{disc}$) appears to show a power-law correlation 
with the inner temperature ($T_{in}$): $L_{\mathrm{disc}}\propto T^{5.3\pm 1.1}$ (see Figure~\ref{figlt}).
In this fit we neglected the observation with kT$_{cor}=1.7$ keV and $\tau=10$
because it has a very low statistics and the disc parameters inferred from it are
not well constrained.

The results obtained modelling Comptonization with the \textit{eqtherm} model
are similar to those obtained with the \textit{comptt} model for both sources. The soft 
component is needed in 5 spectra for X-1 and 7 for X-2. Also in this case X-2 displays 
a dependence of the disc luminosity on the disc inner temperature, that follows the 
relation $L_{\mathrm{disc}}\propto T^{2.1\pm0.5}$.

We analyzed also the RGS data of X-1 and X-2 in an attempt to use them for determining 
the chemical abundances in the local source environment. 
As the sources are faint, only the last observation (the longest with 122 ks)
reached a reasonable statistic for the brightest ULX (X-1), although the RGS 
net count rate is quite low ($1.5\cdot10^{-2}$ count s$^{-1}$). 
The continuum was fitted with the values inferred from the EPIC fit. 
We identify in the spectrum 2 lines in absorption, associated to the O I (535 keV) 
and Fe I (0.709 keV). There may be also two other lines in emission at the
characteristic energy of O VIII K$_{\alpha}$ (0.653 KeV) and Si K$_{\alpha}$ (1.748 KeV), 
but their significance is very low.

The lines do not allow to trace the chemical composition. Therefore we substituted
the \textit{tbabs} model with \textit{tbvarabs}, that allows to vary the  
chemical abundances and grain composition. We determined the abundances fitting the
EPIC continuum and then fixing the abundance of a certain element to zero. The RGS
spectrum was then fitted with this model plus an absorption edge. The parameters
of the edge are then used to compute the abundance.
We did this with the Oxygen line at 0.538 keV and found an abundance consistent with
solar. This result is confirmed repeating the analysis with the EPIC spectra, using 
the same method. 

We performed also a similar abundance analysis on the EPIC spectra of the last observation 
of X-2, finding a sub-solar abundance ($\sim 0.5\pm 0.1$ Z$_\odot$) for Oxygen while the Iron abundance 
is consistent with zero. This is in agreement with the abundances inferred from the HII regions 
in NGC 1313 (e.g. Pilyugin 2001; Hadfield \& Crowther 2007) and from the bubble nebula around X-2 (Ripamonti et al., these Proceedings).

\section{Discussion}
\label{off}

The analysis of all the {\it XMM-Newton} observations shows that the spectra of
X-1 and X-2 can be well reproduced by a Comptonization model plus a soft component.
X-2 appears to show a well defined spectral behaviour in 
the optical depth ($\tau$) versus corona temperature ($T_{cor}$) plane that, apart possibly for one
observation, correlates with the presence of a soft component.

As already suggested by Gladstone et al. (2009),
it is likely that X-1 and X-2 are in different regimes. X-1 shows higher 
average isotropic luminosity ($\sim 10^{40}$ erg s$^{-1}$) and smaller variability during the 
six years of observation. From time to time it shows the presence of a soft component
that emits a significant fraction of the total flux and does not correlate with the 
disc temperature.
As Gladstone et al. (2009) proposed, X-1 could be in the ultraluminous regime, accreting at 
super-Eddington rates and launching powerful winds from the disc/corona.
The accretion disc may be there, but covered by the wind/corona, while the soft component 
may actually represent emission from the wind itself.

On the other hand, X-2 could be in a different regime, in which the average accretion 
rate is at around the Eddington limit. The disk may be still partly visible. 
This hypothesis is enforced by the correlation that we found 
between the disc luminosity and inner temperature, consistent with that of 
a standard disc ($L \propto T^4$) when adopting the \textit{comptt} model. 
The correlation disappears using a simple power-law to describe the spectrum
of the corona, as found by Feng \& Kaaret (2006), because it does not 
take into account the spectral curvature at high energies. 
The corona may become more expanded in the low-luminosity no-disc state, when it covers
the entire disc and no disc component is visible in the spectrum. Clearly, as the corona is always optically
thick, it is not possible to use the disc parameters for estimating the BH mass. 
We found also that for X-2 the corona temperature decreases and the optical depth increases
as the total luminosity goes up, in agreement with the recent analysis of Vierdayanti 
et al. (2010) for Ho IX X-1. So X-2 could be in a state similar to that of Ho IX X-1,
with the corona becoming progressively mass loaded as the luminosity increases (Gladstone et al. 2009).

A couple of ULXs show periodic intensity variations in X-rays which are considered as 
signatures of the orbital period (Kaaret et al. 2006ab; Kaaret \& Feng 2007; Strohmayer 2009). 
X-ray spectra seem to change regularly with orbital phase. It is
possible that these phase related variations may affect any result obtained from
snapshot observations such as presented here. This can be addressed only through a dedicated
X-ray monitoring programme for X-1 and X-2.

We used the RGS high spectral resolution to attempt an estimate of the metallicities of the 
local environments of X-1 and X-2. Because of the low signal-to-noise ratio, 
only the last, longest observation of X-1 could be used. 
The analysis was performed also on the EPIC spectrum. The metallicity in the X-1 environment 
is consistent with being solar and appears to be higher than the average metallicity of NGC 1313. 
The last EPIC spectrum of X-2 was also analyzed with the same method and suggests sub-solar
metallicity. At present we are working at stacking together the RGS spectra of the longest
{\it XMM-Newton} observations (performed in the same state for X-2), trying to improve the signal-to-noise.

\acknowledgements
We thank the referee for useful comments.
We acknowledge financial support through INAF grant PRIN-2007-26.

\bibliographystyle{an} 
\bibliography{/home/pintore/Desktop/relitto/my.bib}

\begin{thebibliography}{}
\bibitem{a} Coppi, P.S.: 2001, MNRAS
\bibitem{b} Done, C., Kubota, A.: 2006, MNRAS 371, 1216
\bibitem{c} Dickey, J.M., Lockman, F.J.: 1990, ARA\&A 28, 215
\bibitem{d} Feng, H., Kaaret, P.: 2005, ApJ 633, 1052
\bibitem{e} Feng, H., Kaaret, P.: 2006, ApJ 650, 75
\bibitem{f} Gladstone, J.C., Roberts, T.P., Done, C.: 2009, ApJ 397, 1836
\bibitem{g} Goncalves, A.C., Soria, R.: 2006, MNRAS 371, 673
\bibitem{k1} Kaaret, P., Simet, M. G., Lang, C. C.: 2006a, Science 311, 491
\bibitem{k2} Kaaret, P., Simet, M. G., Lang, C. C.: 2006b, ApJ 646, 174
\bibitem{k3} Kaaret, P., Feng, H.: 2007, ApJ 669, 106
\bibitem{h} Miller, J. M., Fabbiano, G., Miller, M. C., Fabian, A. C.: 2003, ApJ 585, 37
\bibitem{i} Miller, J. M., Fabian, A. C., Miller, M. C.: 2004, ApJ 607, 931
\bibitem[{Mucciarelli et al. 2007}]{mucciarelli07} Mucciarelli, P. et al.: 2007, ApJ 658, 999
\bibitem{l} Pilyugin L.S.: 2001, A\&A 369, 594
\bibitem{m} Hadfield L. J., Crowther P.A.: 2007, MNRAS 381, 418
\bibitem{n} Stobbart A. M., Roberts T. P.,Wilms J.: 2006, MNRAS 368, 397
\bibitem{s} Strohmayer, T. E.: 2009, ApJ 706, L210
\bibitem{o} Titarchuk L.: 1994, ApJ 434, 570
\bibitem[{Tully 1988}]{tully88} Tully, R. B.: 1988, Nearby Galaxies Catalog. Cambridge Univ. Press, Cambridge
\bibitem{q} Vierdayanti, K., Done, C., Roberts, T. P., Mineshige, S.: 2010, MNRAS 403, 1206
\end{thebibliography}

\end{document}